# Cross-sections for neutral atoms and molecules collisions with charged spherical nanoparticle

M. N. Shneider

*Mechanical and Aerospace Engineering Department, Princeton University, Princeton, New Jersey 08544, USA*
e-mail: m.n.shneider@gmail.com

The paper presents cross sections for collisions of neutral atoms/molecules with a charged nanoparticle, which is the source of the dipole potential. The accuracy of the orbital limited motion (OLM) approximation is estimated. It is shown that simple analytical formulas for the atoms/molecules and heat fluxes, obtained in the OLM approximation, give an error of not more than 15%, and are applicable in all reasonable range of nanoparticles and weakly ionized plasma parameters.

In [1] it was shown that in weakly ionized plasma the flux of the neutral atoms or molecules to the nanoparticle increases noticeably, due to the dipole forces in the vicinity of nanoparticles charged to the floating potential $\varphi_s \sim -T_e/e$, where $T_e$ is the election temperature in plasma. To estimate fluxes of gas particles and heat into spherical nanoparticles of radius $a$, the cross section

$$\sigma_{d,OLM}(v) = \pi a^2 (1 + |U_d(a)|/K), \tag{1}$$

was used in [1], which is similar to the cross section for the ion in the orbit limited motion (OLM) approximation in a dusty plasma theory [2]. Here $U_d(a)$ is the dipole potential on the surface of the nanoparticle, and $K = \dfrac{Mv^2}{2}$ is the initial unperturbed kinetic energy of the atoms/molecules, which are moving to the nanoparticle. Thus, the dipole potential at a given distance *r* from the nanoparticle's surface is given by:

$$U_d(r) = -\frac{1}{2}\alpha E^2 = -\frac{|U_d(a)|a^4}{r^4}, \tag{2}$$

where $\alpha$ is the polarizability of atoms/molecules, and $E(r) = \dfrac{\varphi_s a}{r^2}$ is the electric field in the vicinity of the nanoparticle. The value of the dipole potential is limited by the maximum charge, acquired by nanoparticles in plasma, at which the floating potential is less than the affinity energy of the electron to the surface [3]. For example, according to estimates for a weakly-ionized plasma in helium-carbon mixture, the corresponding values of the dipole potential to all neutral plasma components (helium atoms and carbon atoms and molecules) [1] is given by $|U_d(a)|/kT \leq 2$, where *T* is the translational temperature of the neutral components and ions.

The motion of the atoms/molecules colliding with charged nanoparticles can be considered as collisionless in its vicinity, since the field in the Debye layer is small as compared with the field close to the surface of the nanoparticles [1], and has no appreciable effect on the motion of the neutral particles, for which the mean free path $l_n \gg a$. This condition is satisfied for any neutral gas component in weakly ionized plasma at pressures up to 1 atm for nanoparticles of size $a \sim 1-10$ nm. Using the cross section in (1) allows obtaining simple and clear analytical formulas for the atomic and molecular fluxes and the corresponding heat flux to the nanoparticles in weakly ionized plasma [1]:

$$\Gamma_{d,OLM} = \frac{1}{4} N\bar{v}\left(1 + |U_d(a)|/kT\right), \tag{3}$$

$$H_{d,OLM} = \frac{1}{3} N\bar{v}\bar{\varepsilon}\left[1 + \frac{|U_d(a)|}{kT} + \frac{1}{2}\left(\frac{|U_d(a)|}{kT}\right)^2\right], \tag{4}$$

where $N$, $\bar{v} = \left(\frac{8kT}{\pi M}\right)^{1/2}$, and $\bar{\varepsilon} = \frac{3}{2}kT$ are, correspondingly, the density, average thermal velocity and the average energy of the translational motion of neutral atoms/molecules.

However, strictly speaking, the cross-section in the OLM approximation (1) in all range of the kinetic energies is valid only for particles motion within the potential $U \sim -1/r$ [2,4]. For the considered polarization forces, the dipole potential (2), $U_d(r) \sim -1/r^4$. It is well known (see, for example, [2,4]), that the cross section of collisions with the "source" of the attracting potential $U \sim -1/r^n, n \geq 2$ is limited to relatively slow particles. In this case, the cross section (1) becomes invalid. Naturally, the question arises: how accounting for the correct cross section at low kinetic energies of incident atoms/molecules affects the results of estimations for atoms/molecules and heat fluxes, obtained in [1].

Following [2,4], the motion of the polarized atoms/molecules with an initial kinetic energy $K$, in the vicinity of the charged nanoparticle can be described by the "effective potential"

$$U_{eff} = \frac{\rho^2}{r^2} - \frac{|U_d(a)|a^4}{r^4 K}, \tag{5}$$

where $\rho$ is the impact parameter. For a given $\rho$ the value $r = r_{min}$, which satisfies the condition $U_{eff}(r_{min}, \rho) = 1$, corresponds to the minimum approach of atoms/molecules to the nanoparticle. For slow enough gas particles, moving in the potential $U \sim -1/r^n$, $n \geq 2$, the equation $U_{eff}(r_{min}, \rho) = 1$ can have more than one solution. In this case, the distance of the closest approach $r_{min}$ corresponds to the larger of the two solutions.

Fig. 1 shows examples of solutions to the equation $U_{eff}(r, \rho) = 1$ depending on different values $|U_d(a)|/K$. It is seen that for the fast atoms/molecules, $|U_d(a)|/K \leq 1$, the impact parameter $\rho$ changes monotonically, and for $r_{min} = a$, the solution for the impact parameter is

$$\rho = a(1 + |U_d(a)|/K)^{1/2}. \tag{6}$$

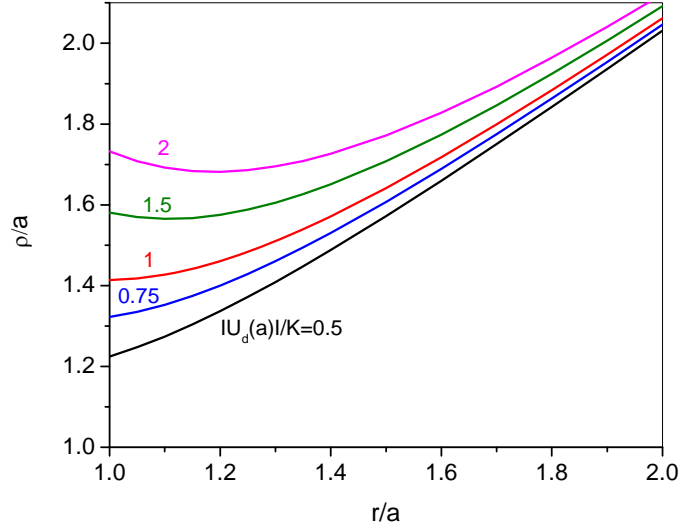

**Fig. 1.** Solutions of the equation $U_{eff}(r,\rho)=1$ for different values $U_d(a)/K$.

The corresponding cross section $\sigma_d(v) = \pi\rho^2$ is the same as (1), which is used in the OLM approximation. On the other hand, for the relatively slow particles, for which $|U_d(a)|/K > 1$ is valid, the solution $U_{eff}(r,\rho)=1$ becomes ambiguous. Following [4], where the solution for the motion of a particle under the influence of force $\vec{F} = -\nabla U$ within an arbitrary attracting potential $U \sim -1/r^n$, $n \geq 2$, was analyzed, we find the impact parameter at which the atoms/molecules "fall" into the nanoparticle

$$\rho(v) = a(4|U_d(a)|/K)^{1/4}, \quad |U_d(a)|/K > 1 \tag{7}$$

The corresponding cross section differs from the OLM and equals:

$$\sigma_{d,1}(v) = 2\pi a^2 (|U_d(a)|/K)^{1/2}, \; |U_d(a)|/K > 1. \tag{8}$$

Assuming a Maxwell distribution function of the atoms/molecules in the plasma and using a cross-section $\sigma_d(v) = \begin{cases} 2\pi a^2(|U_d(a)|/K)^{1/2}, & K < |U_d(a)| \\ \pi a^2(1+|U_d(a)|/K). & K \geq |U_d(a)| \end{cases}$, we find the flux of atoms/molecules and the corresponding heat flux to the nanoparticle

$$\Gamma_d = \frac{N}{a^2}\int_0^\infty v^3 \sigma_d(v) f(v) dv, \tag{9}$$

$$H_d = \frac{N}{a^2}\int_0^\infty \left(\frac{Mv^2}{2} + |U_d(a)|\right) v^3 \sigma_d(v) f(v) dv. \tag{10}$$

Fig. 2 shows a comparison at the same conditions of the computed fluxes (9), (10) with the corresponding fluxes (3) and (4), derived in [1] in OLM approximation.

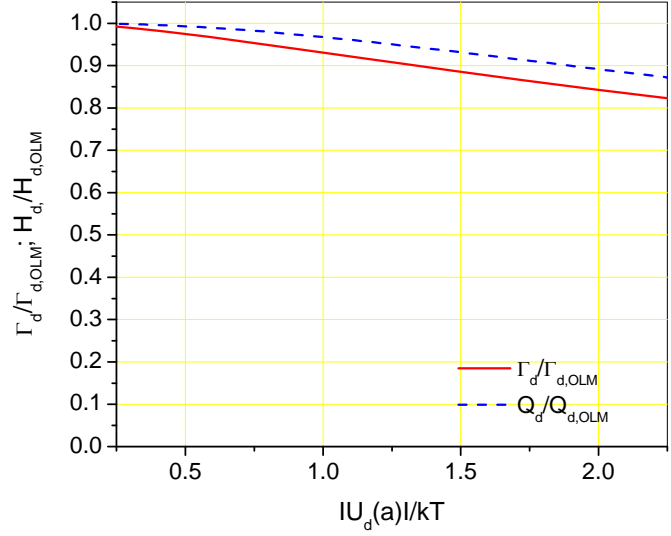

**Fig. 2.** The ratios of corrected atoms/molecules and heat fluxes to the nanoparticle's surface to those obtained in [1] in the OLM approximation.

The performed analysis shows that the OLM approach is quite applicable to describe the fluxes of non-ionized gas particles and related heat fluxes, coming within the dipole potential towards the surface of the charged nanoparticle. Accounting for a more accurate cross-section at low kinetic energies, for the dipole potential $|U_d|/kT \leq 2$, gives a correction not exceeding 15%.

This work was supported by the U.S. Department of Energy, Office of Science, Basic Energy Sciences, Materials Sciences and Engineering Division.